\documentclass[12pt]{iopart}
\usepackage{epsf} 

\begin{document}

\title{Jost function, prime numbers and Riemann zeta function}

\author{Sergio Joffily }
\address{Centro Brasileiro de Pesquisas F\'{\i}sicas,
Laborat\'orio de Cosmologia e F\'isica Experimental de Altas Energias\\   
Rua Dr. Xavier Sigaud, 150,
22290-180, Rio de Janeiro, Brazil }

\date{\today}

\ead{joffily@cbpf.br}

\begin{abstract}
The large complex zeros of the Jost function (poles of the $S$ matrix) in 
the complex wave number-plane for $s$-wave scattering by truncated potentials 
are associated to the distribution of large prime numbers $\{p_n\}$ as well 
as to the asymptotic behavior of the imaginary parts $\{t_n\}$ of the zeros 
of the Riemann zeta function on the critical line. A variant of the Hilbert 
and Polya conjecture is proposed and considerations about the Dirac sea as 
``virtual resonances'' are briefly discussed.

\end{abstract}


There is an old conjecture attributed to 
Hilbert and Polya  about the zeros of the Riemann zeta function 
$\zeta (z)$ on the critical line as eigenvalues of a self-adjoint linear
operator H in some Hilbert space. Ever since Montgomery's \cite{um} discovery 
of these zeros behaving like the eigenvalues of a random hermitian matrix,
attempts have been made in order to find a quantum system with
the Hamiltonian represented by such an operator (see Berry and Keating \cite
{dois}, and references therein). We report in this Letter a variant for the 
above conjecture:
instead of looking for H, whose spectrum coincides with the Riemann zeta
zeros, we are looking for complex momenta poles of the scattering S matrix such
that by a given transformation they are all mapped into the axis $Re\ z=1/2$ 
in coincidence with the Riemann zeroes. The associated quantum system could
be the ``vacuum", interpreted as an infinity of ``virtual resonances",
described by the corresponding S matrix poles.

The first to associate the Riemann hypothesis 
in terms of transient states were Povlov and Fadeev 
\cite{Fadeev} by relating the nontrivial zeros of the zeta function to the
complex poles of the scattering matrix of a particle on a surface of
negative curvature. Here the same problem is discussed in 
the usual scattering by a potential.
We begin by showing that complex momenta zeros of the Jost function for 
s-wave non-relativistic scattering  by repulsive
cutoff potentials, after appropriate transformations, correspond to the
global behavior of heights $\{t_{n}\}$ of the zeros of $\zeta (z)$ on
the critical line, $z_{n}=1/2+i\ t_{n}$, and surprisingly to the global
behavior of the prime sequence $\{p_{n}\}$.  With the aid of these transformations 
it is possible to obtain an approximate formula connecting the $n$th prime
and the $n$th zeta zero, which we believe will be useful to strategies for
primality. The local behavior, 
defined as deviations from the average density of the
zeta zeros, are not obtained by
the potential considered here. This will be the object of
a forthcoming work where a statistical hypothesis with respect to some
residual interaction will be introduced \cite{nove}. Actually 
fluctuations in $\{t_{n}\}$ are interesting for their universality, 
being observed in quantal spectra in different physical systems (see Mehta
 \cite{dez}), and by the connection with chaotic dynamics, (for a review see 
Bohigas \cite{onze}). 

The distribution of primes, $\{p_{n}\}$, among all natural numbers, $n$, in
spite of their local deviation of any known order when viewed at large
possesses regularities that can be approximated by some formulas. 
The approximate number of primes $\pi (x)$ less than a given 
$x$, also called the prime counting function, is given by the prime number
theorem $\pi (x)\sim x/\ln x $ (see Titchmarsh \cite{doze}, Chapter- III), 
where $\ln x$ is the natural logarithm of $x$. 
This relation gives the asymptotic
approximation for $n$th prime $p_{n}$, 
\begin{equation}\label{primo}
p_{n} \sim n\ln \ n\quad ,\qquad \mbox{as}\ n\rightarrow \infty \ .
\end{equation}

The connection between the distribution of prime numbers $\pi (x)$ and 
complex zeros of the zeta function, a kind of duality between the continuum
and discrete in number theory, started with Riemann's 1859 paper (see Edwards
 \cite{treze}, p.299) by introducing methods of analytic function
into number theory. Riemann's zeta function is defined (Ref.\cite{doze}, p.1)
either by the Dirichlet series or by the Euler product 
\begin{equation}\label{zeta}
\zeta (z)=\sum_{n}n^{-z}=\prod_{p}(1-p^{-z})^{-1}\ ,\qquad Rez>1\ ,
\end{equation}
where n runs through all integers and p runs over all primes. $\zeta (z)$ 
can be analytically continued to the whole complex plane, 
except at $z=1$ where it has a simple pole with residue 1. It
satisfies the functional equation 
$\zeta (z)=2^{z}\pi ^{z-1}\sin (\pi z/2)\Gamma (1-z)\zeta (1-z)\ $,
called the reflection formula, where $\Gamma (z)$ is the gamma function. It
is known that $\zeta (z)$ has simple zeros at points $z=-2n,\
n=1,2,\dots $, which are called trivial zeros, with an infinity of
complex zeros lying in the strip $0 < Re\ z < 1$. From the reflection
formula they are symmetrically situated with respect to axis $Re\ z=1/2$, 
and since $\zeta (z^{*})=\zeta ^{*}(z)$, they are also symmetric about
the real axis, so, it suffices to consider the zeros in the upper half of 
strip $1/2\leq Re\ z < 1$. It is possible to enumerate these
complex zeros as $ z_{n}=s_{n}+i\ t_{n}$ , with $ t_{1}\leq
t_{2}\leq t_{3}\leq \dots$, and the following result can be proven
(see Titchmarsh \cite{doze} , p.214) 
\begin{equation}\label{zass}
|z_{n}|\sim t_{n}\sim \frac{2\pi n}{\ln n}\ ,\qquad \mbox{as}\ n\rightarrow \infty .
\end{equation}
The Riemann hypothesis is the conjecture, not yet proven, that all 
complex zeros of $\zeta (z)$ lie on axis $Re\ z=1/2$, called the
``critical line''. Based on this conjecture Riemann improved on Gauss's 
suggestion that $\pi (x)$ approximate the logarithmic integral as 
$x\rightarrow \infty $ with a new prime number formula, taking into account 
local prime fluctuations in terms of nontrivial zeta zeros 
(Ref.\cite{treze}, p.299). 

The Jost function \cite{quatorze} has played a central role in the
development of the analytic properties of the scattering amplitudes. 
In order to recall its
properties, let us consider the scattering of a
non-relativistic particle, without spin, of mass $m$ by a spherically
symmetric local potential, $V(r)$, everywhere finite, behaving at infinity as 
\begin{equation}\label{vass}
V(r)=O(r^{-1-\epsilon }),\quad \epsilon >0,\quad r\rightarrow \infty \ .
\end{equation}
The Jost functions $f_{\pm }(k)$ are defined (see Newton \cite{quinze}, p.341) as the
Wronskian $W$, $f_{\pm }(k)=W[f_{\pm }(k,r),\ \varphi (k,r)]$, 
where $\varphi (k,r)$ is the regular solution of the radial Schr\"{o}dinger
equation 
\begin{equation}\label{schr}
{\lbrack }\frac{d^{2}}{dr^{2}}+k^{2}-V(r){\rbrack }\ \varphi (k,r)=0\ ,
\end{equation}
(in units for which $\hbar=2m=1$) $k$ being 
the wave number and the Jost solutions, $f_{\pm }(k,r)$, are two linearly independent
solutions of equation (\ref{schr}). They satisfy the boundary conditions 
$\lim_{r{\large \rightarrow \infty }} [e^{{\mp }ikr}f_{\pm }(k,r)]=1$,
corresponding to incoming and outgoing waves of unit amplitude.

The properties of the solution of differential equation (\ref{schr}) 
define the domain of analyticity of the Jost
functions $f_{\pm }(k)$ on the complex $k$-plane as well as its symmetry
properties, such as for real potentials, $f_{+}^{*}(k^{*})=f_{-}(k)$. 
The phase of the Jost function is just
minus the scattering phase shift $\delta (k)$, that is 
$f_{\pm }(k)=\vert f_{\pm }(k)\vert \ \ e^{ {\mp }i\delta (k)}\ ,$ 
so that the usual S matrix is given by 
\begin{equation}\label{matrixs}
S(k)\equiv e^{2i\delta (k)}=\frac{f_{-}(k)}{f_{+}(k)}\ .
\end{equation}
The complex poles ($Re\ k \neq 0$) of $S(k)$, or zeros of $f_{+}(k)$,
correspond to the solutions of the Schr\"{o}dinger equation with purely
outgoing, or incoming, wave boundary conditions. Resonances show up as
complex poles with negative imaginary parts, their complex energies being 
\begin{equation}\label{energia}
k_{n}^{2}=\Xi _{n}-i\ \frac{\Gamma _{n}}{2} ,
\end{equation}
where $\Xi _{n}$ and $\Gamma _{n}$ represent the energy and the width, 
respectively, associated with $n$th resonance state. For small $\Gamma _{n}$, 
resonances appear as long-lived
quasistationary states populated in the scattering process. If the width is
sufficiently broad no resonance effect will be observed, as if the lifetime
of state $1/\Gamma _{n}$ is smaller than the time spent by the particle to
traverse the potential: we will
call this kind of S matrix pole as ``virtual resonances'' throughout.
``Virtual resonances'' like the long lived observed ones, in fact, are represented by pairs 
of symmetrical S matrix poles in the complex k-plane, a capture state pole in 
the third quadrant and the decaying state in the fourth quadrant, since they 
give to the asymptotic solution an incoming growing wave and an outgoing decaying 
wave, respectively, exponential in time \cite{dezesete}. That
could be described by Gamow vectors treated by Bohm and Gadella \cite{vsete}
as pairs of S matrix poles corresponding to decay and growth states. 
Examples of broad resonances are the well known large poles, related basically to the
cutoff in the potential considered to be without any physical significance (see
Nussenzveig \cite{dezoito}, p.178).

Condition (\ref{vass}), which together with differential equation (\ref{schr}) establishes
domains of analyticity for $f_{+}(k)$, is not sufficient to determine the
asymptotic behavior for its zeros; for that we would need more information
about the interaction. Then the potential is set equal to zero for $r\geq
R>0 $, which is the cutoff of the potential at arbitrarily large distances $R$. With
this restriction it can be shown that $f_{+}(k)=0$ is an entire equation of order $1/2$
, and according to Piccard's theorem, has infinitely many roots for
arbitrary values of the potential. Position $k_{n}$
of these zeros determines $f_{+}(k)$ uniquely in the whole complex plane, as a consequence
of Hadmard's theorem, which provides \cite{dezenove} 
$f_{\pm }(k)=e^{\pm ikR}f(0)\prod_{n=1}^{\infty }(1-k/k_{n}) $. 
The asymptotic expansion of $k_{n}$, for large n, after introducing 
dimensionless parameter $\beta =kR$, is given by (Ref.\cite{quinze}, p.362) 
\begin{equation}\label{humblet}
\beta _{n}=n\pi  -i\frac{(\sigma +2)\ln\vert n\vert}{2}+\Or(1)
\end{equation}
where $n=\pm 1,\ \pm 2,\ \pm 3\cdots $ and $\sigma $ is to be defined by the
first term of the potential asymptotic expansion, near $r=R$, through $%
V(r)=C(R-r)^{\sigma }+\cdots $, $\sigma \geq 0$ and $r\leq R$.

The connection between the complex zeros of the Jost function and those 
of the Riemann zeta function is obtained by means of the transformation: 
\begin{equation}\label{map}
z=-i\frac{\beta }{2\ Im\ \beta }\ ,
\end{equation}
by which the lower half of complex $\beta $-plane ($Re\ \beta \neq 0$) is
mapped onto the critical axis, $Re\ z=1/2$, of complex $z$-plane. This
suggests a variant for the Hilbert and Polya conjecture, namely, looking for
a potential that gives a Jost function with all zeros on the lower half of 
complex $\beta $-plane ($Re\ \beta \neq 0$) that coincide with 
complex zeros of the Riemannn zeta function after the transformation (\ref{map}). In this
way the Riemann hypothesis follows. For real potentials, these complex $\beta $ 
zeros are located symmetrically about the imaginary axis, then by (\ref{map}) they
will be mapped symmetrically about the real axis into the critical line.

Now we show that for cutoff potentials transformation (\ref{map}) 
gives rise to complex Jost zeros with the same asymptotic behavior as the complex Riemann zeta
zeros, being all in the critical line. If $\{\beta_{n}\}$ are the zeros of $f_{+}(\beta)$ 
then by (\ref{map}) we get 
\begin{equation}\label{linha}
z_{n}= \frac{1}{2}-i\frac{\ Re\ \beta_{n} }{2\ Im\ \beta_{n} }\ ,
\end{equation}
from (\ref{humblet}), with $\sigma =0$, we see that $\{Im\ z_{n}\}$ has the same asymptotic
expansion as $\{t_{n}\},$ given by (\ref{zass}), i.e., 
\begin{equation}\label{linhass}
Im\ z_{n}=\frac{t_{n}}{4}\quad \mbox{as}\quad n\rightarrow \infty \ ,
\end{equation}
which means that for each resonance, ratio $\{2Re\ \beta_{n}/Im\ \beta_{n}\}$ corresponds 
to the height of the zeta zero on the critical line. 

On the other hand, after introducing dimensionless
quantities, energy $E_{n}=R^{2}\Xi _{n}$ and widths $G_{n}=R^{2}\Gamma _{n} $
, equation (\ref{energia}) is written as $\beta_{n} ^{2}=E_{n}-iG_{n}/2$, the 
dimentionless widths $\{G_{n}\}$, defined as 
\begin{equation}\label{width}
G_{n}=4\ Re\beta _{n}\ \ Im\beta _{n}\ ,
\end{equation}
after taking into account (\ref{humblet}), when $\sigma =0$, shows the same
asymptotic expansion for large primes (\ref{primo}), given by the prime number
theorem, 
\begin{equation}\label{widthass}
G_{n}=4\pi p_{n}\quad \mbox{as}\quad n\rightarrow \infty .
\end{equation}
Then $n$th large complex Jost zeros are also related to $n$th large
primes, showing an asymptotic connection between primes and complex Riemann
zeta zeros, in a one-to-one correspondence. 

In the scattering on a surface of constant negative curvature \cite{Fadeev} the 
potential is replaced by imposing that the particle move on the given surface in 
order to obtain the scattering function $S(k)$ in terms of the Riemann zeta 
function. Provided the Riemann hypothesis is true the poles of 
$S$-function in the complex $k$-plane are given by $k_{n} = t_{n}/2-i/4$ 
\cite{Fadeev,Wardlaw}. By using transformation (\ref{map}) we obtain 
$z_{n}=1/2+i\ t_{n}$ in coincidence with the complex zeros of the zeta function 
as expected. It follows also from the usual 
resonance width definition (\ref{energia}) that 
$\Gamma _{n} = t_{n}/2$, is therefore not constant as considered in \cite{Wardlaw}, 
exhibiting the same fluctuation of the height of the zeta zero in the critical line, 
in accordance with the Wardlaw and Jaworski \cite{Wardlaw} results on the time 
delay fluctuation for this kind of unusual scattering.
\begin{figure}[tbh]
\begin{center}
\epsfysize=6.8cm
\epsfbox{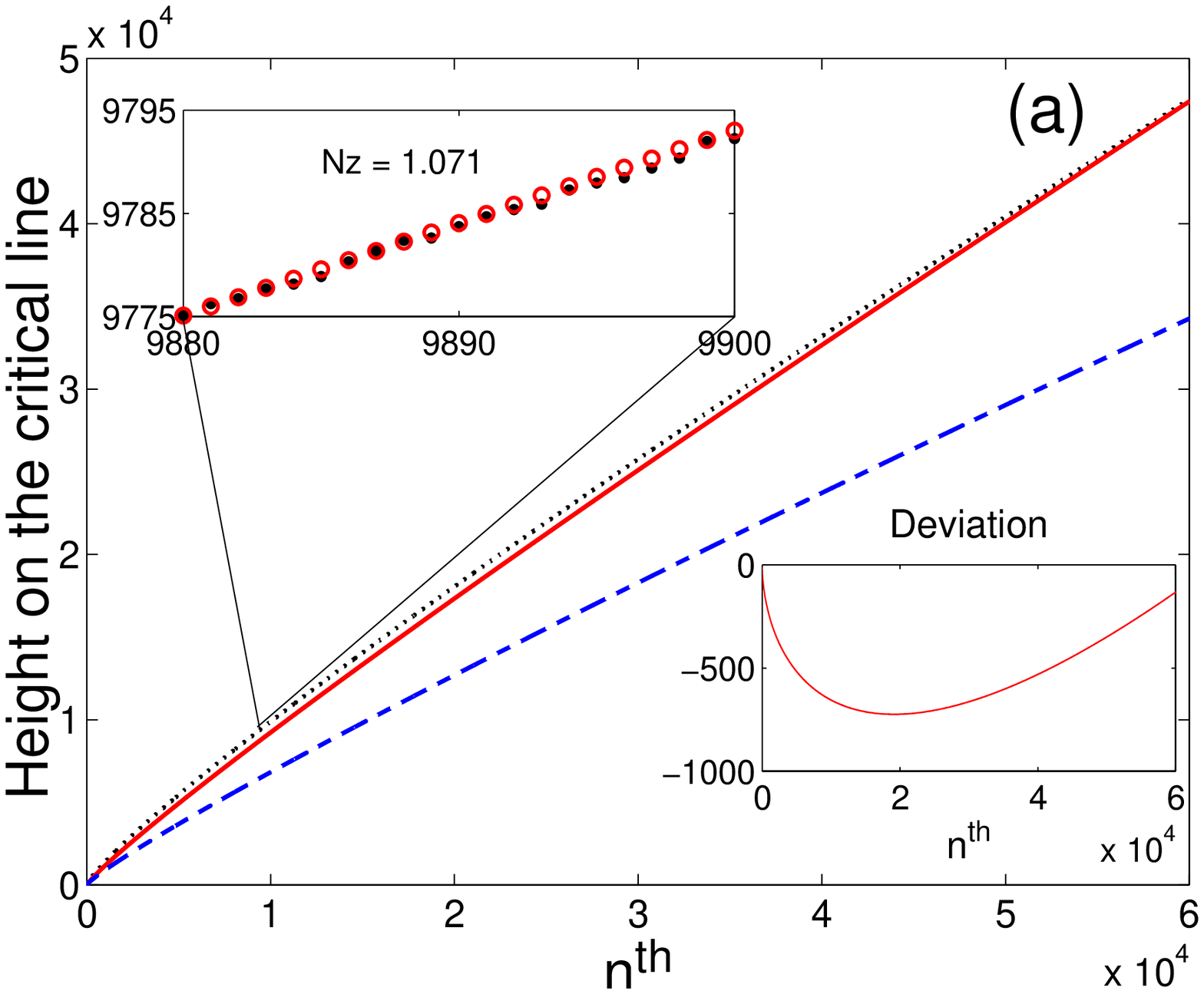}
\end{center}
\begin{center}
\epsfysize=6.8cm
\epsfbox{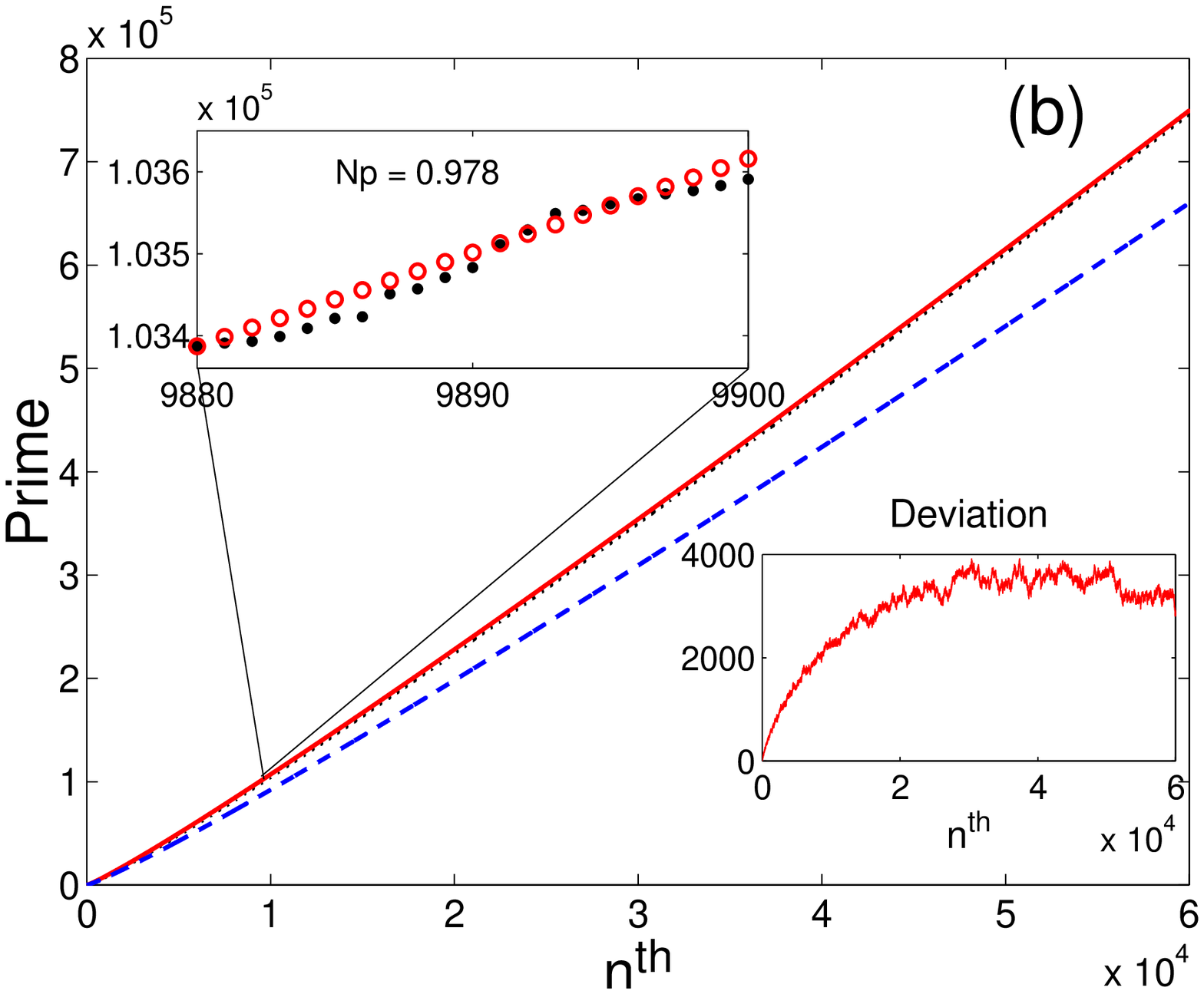}
\end{center}
\caption{(a) The height of nontrivial zeta zeros on the critical line. 
The full line according to
energy/width ratios  $\{4\ \pi E_{n}/G_{n}\}$ from the Jost zeros; 
the dashed line from the asymptotic formula  (\ref{zass}); the points 
are computed by Odlyzko [18]. 
(b) Primes in order of size. The full line 
according to dimensionless width $\{G_{n}/4\pi \}$ from the Jost zeros;
the dashed line according to the prime number theorem (\ref{primo}); 
the dotted line taken from the table of Caldwell [19]. 
Insets are defined in the text.}
\end{figure}

In order to look for physical systems associated to the proposed $S$-function, 
we will return to the conventional scattering. In this connection it is 
interesting to examine more explicitly the potential parameter dependence in 
the above asymptotic expression (\ref{humblet}). Khuri \cite{Khuri} has recently 
proposed a modification to the inverse scattering problem in order to obtain the 
potential whose coupling constant spectrum coincides with the Riemann zeta zeros. 
The model we have chosen is the
non-relativistic s-wave scattering by a spherically symmetric 
barrier potential, $V(r)= V_{0}$ for $ r < R $ and $0$ for otherwise.  
From the stationary scattering solution with this potential one obtains the
Jost function $f_{+}(k)=e^{ikR}[k^{\prime }cos(k^{\prime }R)-ik\sin (k^{\prime }R)]$ ,
where $ k^{\prime }{}^{2}=V_{0}-k^{2}$. 
Introducing dimensionless parameters: $\alpha =k^{\prime }R,\ \beta =kR$ and 
$v=V_{0}R^{2}$, the zeros of the Jost function are given by the solution of
the complex transcendental equation 
\begin{equation}\label{trans}
\sqrt{\beta ^{2}-v}\ \cot \sqrt{\beta ^{2}-v}=i\beta 
\end{equation}
for each value of potential strength $v$. The displacements of the roots, 
$\beta _{n}$, in
the $\beta $-plane with the variation of the potential strength are shown by
Nussenzveig \cite{vdois}. These zeros are all in the lower half of complex $%
\beta $-plane and located symmetrically about the imaginary axis. For finite 
$v\neq 0$, in equation (\ref{trans}), the large values of the $\beta _{n}$ zeros are
asymptotically determined \cite{dezesete,vdois}, as 
\begin{equation}\label{barass}
\beta _{n}=\pm n\pi\  {\mp }\frac{\ln (2\ n\pi /|\sqrt{}v|)}{n\pi}  
 -i(\ln(2\ n\pi /|\sqrt{}v|)\ \mp\frac{1}{n\pi}) \ \ , 
\end{equation}

\noindent n being a large positive integer. This formula gives the dependence
of the potential strength $v$, considered as a free parameter, in the determination of the
asymptotic Jost zeros $\beta _{n}$ to be compared, after the transformation
(\ref{linha}), with the zeta heights $t_{n}$ on the critical line, and after (\ref{width}) 
with asymptotic prime numbers $p_{n}$. By using asymptotic expansion (\ref{barass}) we calculate  
the energy/width ratios $\{4\pi E_{n}/G_{n}\}$, 
where $4\pi E_{n}/G_{n}\sim \pi Re\ \beta_{n}/Im\ \beta_{n}$, 
which are compared to $\{t_{n}\}$ computed by Odlyzko \cite{vquatro} and 
shown in figure 1(a), for $v=2$. 
Then we see the approximate agreement from the beginning, for n up to $6 \times 10^{4}$.
The same Jost zeros $\beta _{n}$, for $v=2$, after
transformation (\ref{width}), give dimensionless widths $\{G_{n}/4\pi \}$ that
are compared with prime sequence $\{p_{n}\}$ taken from the table of Caldwell \cite
{vcinco}, as shown in figure 1(b). The numerical factor $1/4\pi$ was imposed to $G_{n}$ 
by the asymptotic limit (\ref{widthass}) from the prime number theorem. 
Here we see the agreement with the global behavior
of the sequence of primes, from the beginning in the same range. 
In the upper left hand corner in both figures  
the local zeta zero and prime fluctuactions shown by dots are not present when 
obtained by the corresponding Jost zeros (open circles), here normalized on the 
$ 9880 $-th zero; the normalization factor corresponding to the 
height on the critical line is 
$ Nz = 1.071$ and to the prime being $Np = 0.978$. The inset of each
figure in the lower right hand corner shows deviation in the height of zeta zeros 
figure 1(a) and prime figure 1(b), obtained by the corresponding Jost zeros, 
from true values 
taken from tables \cite{vquatro,vcinco} in the range considered.

Finally, a conjecture is made in order to associate the Dirac sea as ``virtual 
resonances". The hole theory is the interpretation of the negative energy solution
of the relativistic single-particle Dirac equation in which the vacuum consists of 
all these negative energy states being filled with electrons, such that it could be
considered from the beginning as a many-particle system being described by the 
formalism of the single-particle theory.
According to Dirac (Ref.\cite{vseis}, p.34), ``The vacuum must be a state with a lot of
particles present corresponding to some stationary solutions of the
Schr\"odinger equation. But there are no known solutions of this Schr\"odinger
equation - not even a solution which could represent the vacuum''. The
question about the vacuum structure was bypassed by the second quantization,
where a vacuum state is assumed and an operator defined, in order to create
an electron when applied to this state, without knowing what the vacuum
state really is. We would like to suggest a description of the vacuum structure 
as being a dynamical system described by ``virtual resonances'', completely
independent of the second quantization. Instead of looking for a stationary
solution, we look for a transient state of very short time duration. Specifically
we propose a dynamical model for the vacuum described by infinite denumbered virtual 
resonances, with discrete widths and energies, which could be useful in the description 
of quantum chaos, which is under investigation. It could perhaps be
verified experimentally by some devise that amplifies the vacuum fluctuation;
the fluctuating nature of the Casimir force was recently discussed by Bartolo et al. 
\cite{vnove}. In this case, the universality of level fluctuation laws of the 
spectra of different quantum systems (nuclei, atoms and molecules)
 \cite{dez,onze} could be understood by the ``vacuum" role as a dissipative system 
\cite{Callen}.
It had already been shown by Maier and Dreizler \cite{voito}, in a Dirac
particle scattering in (1+1) dimensions by an electrostatic square well
potential, that complex momenta poles of the S matrix exhibit the
particle-antiparticle content of the Dirac theory and in the limit of weak
potential strength the poles distribution are similar to the nonrelativistic case. 

In summary, we have shown that the zeros of Riemann's zeta function are related 
to the zeros of the Jost function for cutoff potentials in the complex momenta 
plane in a one-to-one correspondence.
The energy/width ratios of the large ``virtual resonances'' are associated to the 
nontrivial zeta zeros and the corresponding widths related to the prime sequence.
In analogy to the mean field it is expected, by means of a statistical hypothesis, 
that the above relationship would be improved and the distribution of the virtual 
resonances would reflect the chaotic nature of the vacuum.
\ack
We thank C. A. Garcia Canal, M. Novello, J. M. Salim, B. Schroer and I. D. Soares, 
for their interest in the work and useful conversations.

\end{document}